\documentclass[5p,twocolumn,number,times]{elsarticle}
\usepackage{amsmath}
\usepackage{amssymb}
\usepackage{graphicx}
\usepackage{hyperref}
\usepackage[english]{babel}
\usepackage{babel}
\usepackage[usenames,dvipsnames]{color}

\input{tcilatex}
\begin{document}

\begin{frontmatter}

\title{Metrological power of single-qubit dynamical Casimir effect in circuit QED}
\author[if]{A.P. Costa}
\author[if]{H.R. Schelb}
\author[if,cif]{A.V. Dodonov\corref{cor1}}
\ead{adodonov@unb.br}
\cortext[cor1]{Corresponding author. Tel.: +55 61 31076076; Fax: +55 61 31077712}
\address[if]{Instituto de F\'{\i}sica, Universidade de Bras\'{\i}lia, Caixa Postal 04455, CEP
70910-900, Bras\'{\i}lia, DF, Brasil}
\address[cif]{International Center of Physics, Institute of Physics, University of Brasilia, 70910-900, Brasilia, DF, Brazil}

\begin{abstract}
We consider a nonstationary circuit QED system described by the quantum Rabi model, in which an artificial two-level atom with a tunable transition frequency is coupled to a single-mode resonator. We focus on regimes where the external modulation takes the form $\sin[\eta(t) t]$, with the modulation frequency $\eta(t)$ varying slowly and linearly in time near $2\nu$ and $4\nu$, $\nu$ being the resonator frequency. Starting from the vacuum state, we numerically compute the Quantum Fisher Information for single-mode phase and displacement estimation, showing that it significantly exceeds the classical limits for the same average photon number, even in the presence of dissipation. Thus, appropriate parametric modulation of the qubit not only simulates the dynamical Casimir effect but also enables the generation of nonclassical states of light that offer a metrological advantage over classical states of equivalent energy.
\end{abstract}


\end{frontmatter}{}

\section{Introduction}

The most fundamental problem in metrology is the estimation of an unknown physical parameter, such as the amplitude of a DC or AC electromagnetic field, a small mechanical force, tiny displacements, coupling constants, etc \cite{a,b,c,zub,lloyd,d,k,z1,z2,z3,z4}. Quantum metrology exploits quantum mechanical principles to enhance the precision of parameter estimation beyond classical limits (see reviews \cite{re1,re2,avs,Sidhu} and references therein). It plays a crucial role in improving time and frequency standards, detecting gravitational waves, performing sensitive optical phase measurements, synchronizing clocks, conducting interferometry with interacting systems, magnetometry, and other applications \cite{a,b,c,avs,Sidhu,jiang}. Measurement devices that use quantum systems in nonclassical states as “probes” can outperform their classical counterparts without consuming additional resources, where resources typically refer to the number of photons or atoms \cite{avs}.

At the core of quantum metrology lies the concept of Quantum Fisher Information (QFI), a fundamental quantity that sets the ultimate sensitivity limit of any unbiased estimator through the quantum Cramér–Rao inequality \cite{re1,re2,avs}. The quantum Cramér–Rao bound depends solely on the density operator of the probe and the unknown parameter to be estimated, and it can, in principle, always be saturated by an appropriate measurement. Metrological power refers to any quantum enhancement arising from the nonclassicality of the probe state \cite{zub,avs}, and QFI has become an essential tool for characterizing quantum states in terms of their sensitivity to external perturbations, thereby quantifying their metrological utility under both ideal and noisy conditions. In particular, Gaussian states of the electromagnetic field have found numerous applications in quantum metrology \cite{Bagan,Pinel,Bina,jiang,Safranek}, with the squeezed vacuum state playing a prominent role as the most sensitive pure Gaussian state for phase estimation (for a fixed average photon number) \cite{Monras}. Furthermore, it has recently been shown that the QFI for quadrature-displacement sensing uniquely determines the maximum achievable precision enhancement in both local and distributed quantum metrology using any linear or nonlinear single-mode unitary transformation, providing a unified measure of metrological power \cite{zub}.

In this Letter, we numerically study the QFI for non-Gaussian mixed states of the electromagnetic field generated from vacuum in a nonstationary circuit Quantum Electrodynamics (circuit QED) setup through parametric modulation of the qubit transition frequency. Circuit QED is one of the most promising platforms for quantum technologies, including quantum computing and quantum simulation, due to its high level of controllability and scalability. In circuit QED, superconducting qubits and resonators can be manipulated with unprecedented precision, while dissipative losses remain much smaller than the atom–field coupling strengths \cite{dir4,j1,j2,j3,j4,jer,c1,c4,c5}. Notably, real-time control over the effective resonator frequency via an external magnetic field has enabled experimental implementations of analogs of the dynamical Casimir effect (DCE) \cite{Wilson11-Nature,Paraoanu13}, i.e., the generation of photons from vacuum due to external variations of boundary conditions or material properties of the intracavity medium \cite{rev4,rev5}. The nonclassical properties of field states generated in circuit DCE were studied in \cite{disc,Johan13}, while applications of thermal two-mode squeezed states for quantum metrology were analyzed in \cite{wilkins}.

The controllability of qubit parameters also allows DCE analogs using a single qubit as the source of system perturbation \cite{simone,AVD09,ibe,AVD14}. Here, we consider sinusoidal modulation of the qubit’s level splitting with a linearly time-varying modulation frequency, as originally proposed in \cite{AVD16}. We focus on modulation frequencies near $2\nu$ and $4\nu$, where $\nu$ is the resonator frequency, leading to photon creation from vacuum via two- and four-photon transitions, respectively \cite{AVD19}. We model the system using the standard Markovian master equation of Quantum Optics with a time-dependent Rabi Hamiltonian and numerically study the evolution of the QFI for phase \cite{Monras,Bagan} and phase-space displacement \cite{d,dis} estimations, as well as the average photon number, its variance, qubit excitation, resonator field purity, and negativity. Our main finding is that, for appropriately chosen system parameters, one can generate mixed states of light with high QFI values unattainable by classical field states with the same energy. Moreover, in some cases, these field states exhibit higher QFI values than the squeezed vacuum state for the same average photon number. To our knowledge, this is the first demonstration that single-qubit DCE can be used to produce nonclassical states of light with applications in quantum metrology.

\section{Mathematical formalism}

We consider a two-level artificial atom (qubit) in
a superconducting resonator of frequency $\nu $. The system is described by
the quantum Rabi Hamiltonian \cite{larson,solano,braak,xie} ($\hbar =1$)%
\begin{equation}
H=\frac{\Omega \left( t\right) }{2}\sigma _{z}+\nu n+g\left( a+a^{\dagger
}\right) \left( \sigma _{+}+\sigma _{-}\right) \,,  \label{QRM}
\end{equation}%
where $a$ and $a^{\dagger }$ are the annihilation and creation operators of
the electromagnetic field, $n=a^{\dagger }a$ is the photon number operator
and $g$ is the light--matter coupling constant \cite{ultra,ultra2,celso}.
The atomic operators are $\sigma _{z}=|e\rangle \langle e|-|g\rangle \langle
g|$, $\sigma _{+}=|e\rangle \langle g|$ and $\sigma _{-}=\sigma
_{+}^{\dagger }=|g\rangle \langle e|$, where $|g\rangle $ and $|e\rangle $
denote the ground and the excited states, respectively. To induce the photon
generation from vacuum, we suppose that the atomic transition frequency is
modulated externally as%
\begin{equation}
\Omega (t)=\Omega _{0}+\varepsilon \sin [\eta (t)t]~,
\end{equation}%
where $\Omega _{0}$ is the bare qubit frequency, $\varepsilon \ll \Omega
_{0} $ is the modulation amplitude and $\eta (t)$ is the time-dependent
modulation frequency. Following the proposal \cite{AVD16}, we consider a
slow linear sweep, $\eta (t)=\eta _{0}+\alpha t$, where $\eta _{0}$ and $%
\alpha $ are constant parameters. We take into account the effects of weak
Markovian dissipation \cite{jer,veloso} via the the standard master
equation of Quantum Optics, also known as Gorini-Kossakowski-Sudarshan-Lindblad (GKSL) master equation,%
\begin{equation}
\frac{\partial \rho }{\partial t}=-i\left[ H,\rho \right] +\gamma \mathcal{D}%
[\sigma _{-}]\rho +\gamma _{\phi }\mathcal{D}[\sigma _{z}]\rho +\kappa
\mathcal{D}[a]\rho ,  \label{fun}
\end{equation}%
where $\rho $ is the total system density operator, $\gamma $ ($\gamma
_{\phi }$) is the atomic damping (pure dephasing) rate, $\kappa $ is the
cavity damping rate, $\mathcal{D}[\sigma ]\rho \equiv \sigma \rho \sigma
^{\dagger }-\left( \sigma ^{\dagger }\sigma \rho +\rho \sigma ^{\dagger
}\sigma \right) /2$ is the Lindblad superoperator and we assumed zero
temperature reservoirs.

We now briefly review the concept of quantum Fisher information, whose properties are discussed in detail in the reviews \cite{avs,re1,re2,Sidhu}. To estimate an unknown parameter $\theta$, one collects measurement outcomes from experiments and constructs an estimator $\hat{\theta}$, a function of the data that provides the best guess for $\theta$. Common examples of parameter estimation using bosonic fields include the estimation of phase and phase-space displacements. For an unbiased estimator, in the limit of a large number $n$ of measurements, the precision is bounded from below by the Cramér–Rao bound: $\Delta ^{2}\theta \geq %
\left[ nF\left( \theta \right) \right] ^{-1}$. Here
\begin{equation}
F\left( \theta \right) =\int dxp\left( x|\theta \right) \left[ \frac{%
\partial }{\partial \theta }\ln p\left( x|\theta \right) \right] ^{2}
\label{eq1}
\end{equation}%
is the Fisher information of the probability distribution $p(x|\theta)$, which is the conditional probability of obtaining measurement outcome $x$ given the parameter value $\theta$. Higher precision can be achieved by increasing the number of measurements and/or the Fisher information $F(\theta)$.

The QFI, $\mathcal{F}(\rho,\theta)$, is the supremum of the Fisher information over all possible quantum measurements described by positive operator-valued measures (POVMs) \cite{avs,Sidhu,re2}, where $\rho$ denotes the density operator of the system used as the probe. In this sense, QFI quantifies the maximum information about $\theta$ that can be imprinted onto the probe state. The ultimate limit on extracting information about an unknown parameter $\theta$ via a quantum measurement is given by the quantum Cramér–Rao bound, $\Delta ^{2}\theta \geq \left[ n\mathcal{F}\left( \rho
,\theta \right) \right] ^{-1}$, which depends only on the probe’s density operator and the unknown parameter $\theta$. Intuitively, for states that differ slightly in the value of $\theta$, the more distinguishable the states, the more precisely $\theta$ can be estimated. Notably, QFI is independent of the final measurement procedure. Optimizing measurement precision thus involves choosing a probe state with large QFI and a measurement strategy such that Eq. (\ref{eq1}) approaches the QFI. It can be shown \cite{avs} that the quantum Cramér–Rao bound can always be saturated by an appropriate measurement, so $\mathcal{F}(\rho,\theta)$ quantifies the usefulness of a probe $\rho$ for measuring a given $\theta$.

For the \emph{phase estimation} using unitary encoding, the QFI reads \cite{avs}%
\begin{equation}
\mathcal{F}_{ph}=\frac{1}{2}\sum_{i,j}\frac{\left( p_{i}-p_{j}\right) ^{2}}{%
p_{i}+p_{j}}\left\vert \langle i|\hat{n}|j\rangle \right\vert ^{2}\,,
\label{QFI}
\end{equation}%
where $p_i$ and $|i\rangle$ are the eigenvalues and eigenvectors of the reduced cavity density operator, $\rho_{cav}(t) = \mathrm{Tr}_{qubit}[\rho(t)]$. The inequality $\mathcal{F}_{ph} \leq \langle n \rangle_{clas}$ holds for any classical state $\rho_{clas}$, where $\langle n \rangle_{clas} = \mathrm{Tr}(\rho_{clas} a^\dagger a)$ is the corresponding average photon number. Therefore, any quantum state with $\mathcal{F}_{ph} > \langle n \rangle$ exhibits metrological power \cite{avs,Sidhu}. Among pure Gaussian states, the squeezed vacuum state, $\rho_{sv} = \exp\left[(\varepsilon^* a^2 - \varepsilon a^{\dagger 2})/2\right] |0\rangle$ with a complex parameter $\varepsilon$, is the most sensitive state for a given average photon number. Its QFI is $\mathcal{F}_{ph}(\rho_{sv}) = 2(\langle n \rangle_{sv}^2 + \langle n \rangle_{sv})$. So we also calculate numerically the ratio
\begin{equation}
r=\frac{1}{2}\frac{\mathcal{F}_{ph}}{\left\langle n\right\rangle
^{2}+\left\langle n\right\rangle }~,  \label{rar}
\end{equation}%
which compares the QFI of the field state generated via single-qubit DCE with that of the squeezed vacuum state generated in the standard DCE \cite{rev4,rev5}, assuming the same average photon number.

\begin{figure}[tbh]
\begin{center}
\includegraphics[width=0.48\textwidth]{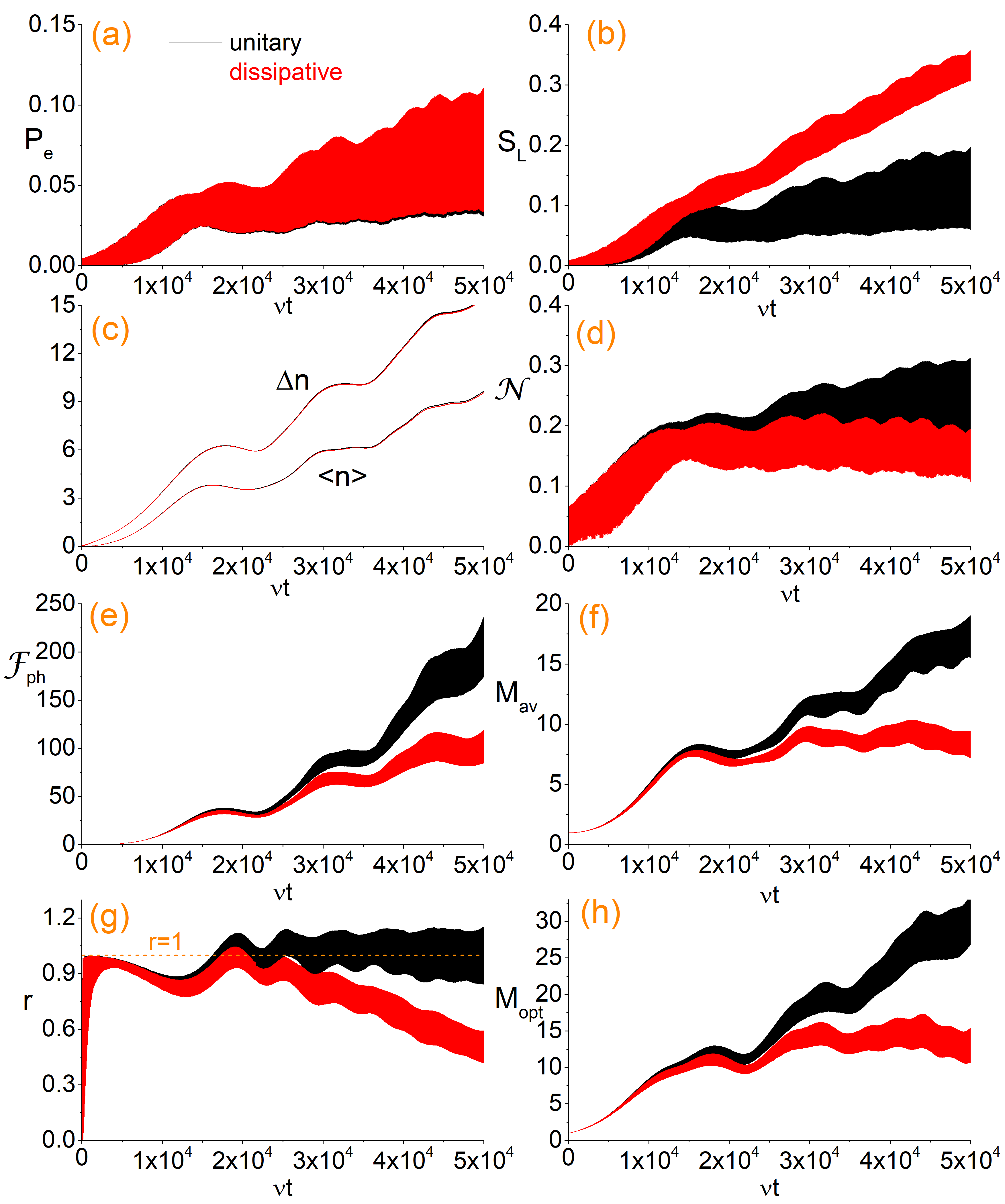} {}
\end{center}
\caption{Behavior of some relevant quantities under unitary (black
curves) and dissipative (red curves) dynamics, obtained by solving
numerically the master equation (\protect\ref{fun}) for the initial state $%
|g,0\rangle $ and the two-photon DCE regime. a) Atomic excitation
probability (the difference between the unitary and dissipative evolutions
is practically invisible). The curves appear broad due to fast oscillations
(with frequencies $\sim \protect\nu $). b) Linear entropy of the cavity
field, which measures entanglement under unitary evolution. c) Average photon
number and the standard deviation. d) Negativity, Eq. (\protect\ref{nee});
nonzero values attest entanglement. e) Quantum Fisher Information for single
mode phase estimation, Eq. (\protect\ref{QFI}). f) $M_{av}$, Eq. (\protect
\ref{Mav}). g) Ratio $r$, Eq. (\protect\ref{rar}). h) $M_{opt}$, Eq. (%
\protect\ref{Mopt}). Parameters: $g=0.05\protect\nu $, $\Omega _{0}=0.5%
\protect\nu $, $\protect\varepsilon =0.08\Omega _{0}$, $\protect\eta %
_{0}=2.00655\protect\nu $, $\protect\alpha =2\times 10^{-8}\protect\nu ^{2}$
and $\protect\lambda =\protect\lambda _{\protect\phi }=\protect\kappa %
=10^{-6}\protect\nu $.}
\label{fig1}
\end{figure}

\begin{figure}[tbh]
\begin{center}
\includegraphics[width=0.48\textwidth]{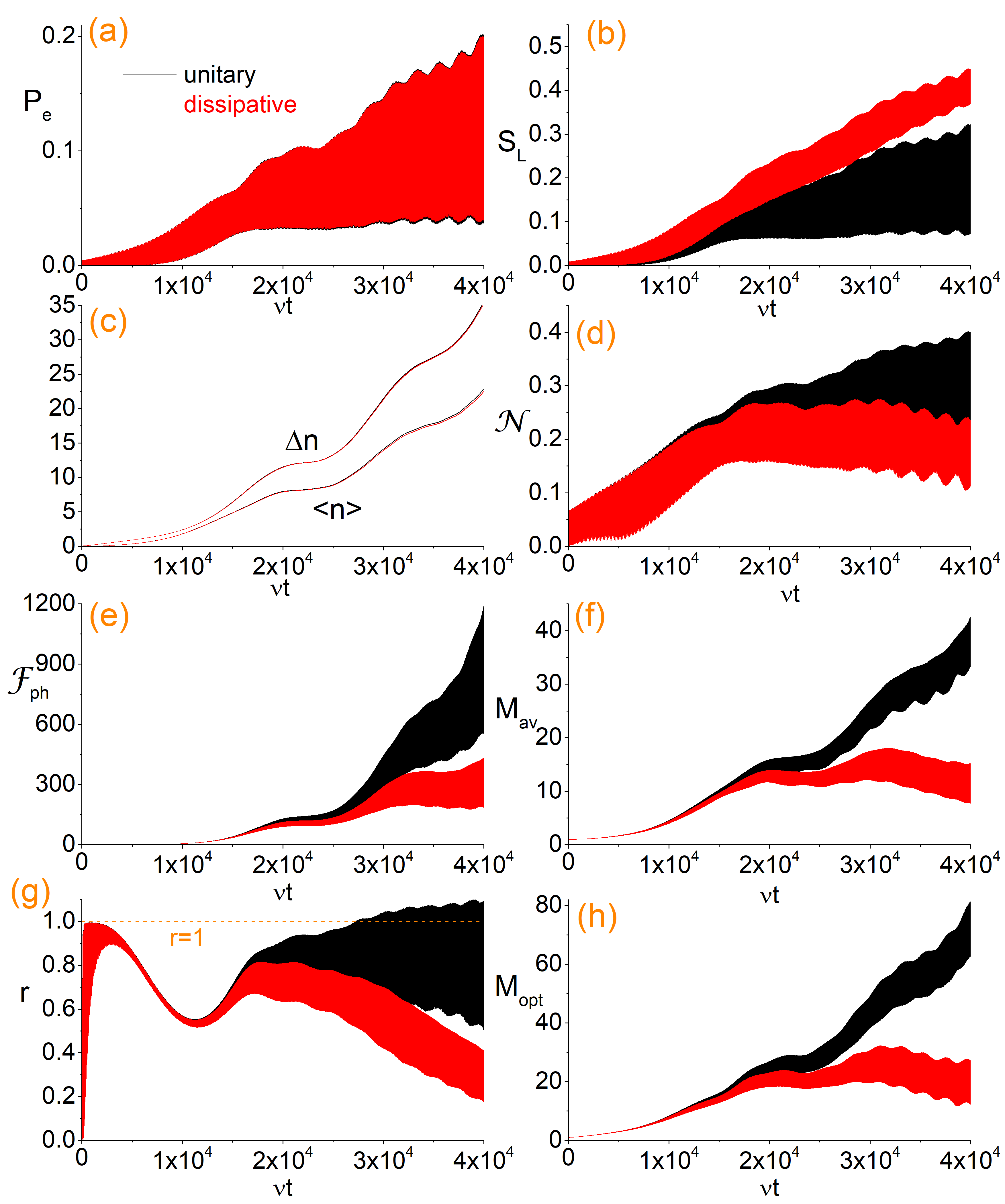} {}
\end{center}
\caption{Similar to Fig. \protect\ref{fig1}, but for slightly different
modulation form: $\protect\eta _{0}=2.00715\protect\nu $ and $\protect\alpha %
=-5\times 10^{-8}\protect\nu ^{2}$. Notice that $\left\langle n\right\rangle
$ and QFI are much higher than in Fig. \protect\ref{fig1}, illustrating how
small changes in the modulation form strongly affect the system dynamics. }
\label{fig2}
\end{figure}

To estimate the \emph{phase-space displacement} of a single-mode field, one must study the $2\times 2$ quantum Fisher
information matrix \cite{avs} with the elements%
\begin{equation}
\left( F_{disp}\right) _{kl}=\sum_{i,j}\frac{\left( p_{i}-p_{j}\right) ^{2}}{%
p_{i}+p_{j}}\langle i|x^{\left( k\right) }|j\rangle \langle j|x^{\left(
l\right) }|i\rangle ~,
\end{equation}%
where $x^{\left( 1\right) }=a+a^{\dagger }$ and $x^{\left( 2\right) }=\left(
a-a^{\dagger }\right) /i$. The \emph{average} Fisher information over all
possible quadrature directions is%
\begin{equation}
M_{av}=\frac{1}{4}\limfunc{Tr}\left( \mathbf{F}_{disp}\right) ~,  \label{Mav}
\end{equation}%
while the \emph{maximum} Fisher information over all possible quadrature
directions reads%
\begin{equation}
M_{opt}=\frac{1}{2}\lambda _{\max }\left( \mathbf{F}_{disp}\right) ~,
\label{Mopt}
\end{equation}%
where $\lambda _{\max }\left( \mathbf{F}_{disp}\right) $ is the maximum
eigenvalue of $\mathbf{F}_{disp}$. For any classical state, the following inequalities
hold: $M_{av}(\rho _{clas}),M_{opt}(\rho _{clas})\leq 1$. Therefore, any
quantum state with $M_{av}(\rho _{cav})>1$ or $M_{opt}(\rho _{cav})>1$ also
offers metrological power for the displacement estimation.

\section{Numeric results}

We solved numerically the master equation (\ref{fun}) for the initial state $%
|g,0\rangle $, using the Runge-Kutta-Verner fifth-order and sixth-order
method. Besides the metrological quantities described in the previous section
($\mathcal{F}_{ph}$, $r$, $M_{av}$ and $M_{opt}$), we calculate the qubit
excitation probability, $P_{e}=\limfunc{Tr}\left[ \rho |e\rangle \langle e|%
\right] $, the average photon number $\left\langle n\right\rangle $, its
standard deviation, $\Delta n=\sqrt{\left\langle
n^{2}\right\rangle -\left\langle n\right\rangle ^{2}}$, the linear entropy of
the cavity field, $S_{L}=1-\limfunc{Tr}\left( \rho _{cav}^{2}\right) $, and
the negativity \cite{negati}
\begin{equation}
\mathcal{N}\left( \rho \right) =\left\vert \sum_{\lambda _{i}<0}\lambda
_{i}\right\vert ~,  \label{nee}
\end{equation}%
where $\lambda _{i}$ are the eigenvalues of the partial transpose of $\rho $
with respect to any subsystem. For any bipartite system, nonzero negativity
implies entanglement. In all the figures presented below, we consider the
modulation amplitude $\varepsilon =0.08\Omega _{0}$ and the realistic
dissipative parameters $\lambda =\lambda _{\phi }=\kappa =10^{-6}\nu $ \cite%
{rates}. For comparison, the system behavior under the unitary dynamics is also
shown; in this case, the linear entropy, beside measuring the purity of the
cavity field, also attests entanglement.

In Figs. \ref{fig1} and \ref{fig2}, we set $g=0.05\nu $ (below the ultrastrong regime \cite{ultra,ultra2}), $\Omega _{0}=0.5\nu
$ and consider a linear sweep of the modulation frequency in the vicinity of
the standard DCE resonance, $\eta \approx 2\nu $. In Fig. \ref{fig1}, we
assume $\eta _{0}=2.00655\nu $ and $\alpha =2\times 10^{-8}\nu ^{2}$, while
in Fig. \ref{fig2}, $\eta _{0}=2.00715\nu $ and $\alpha =-5\times 10^{-8}\nu
^{2}$. In both cases, we see that a few dozen of photons are generated from vacuum,
while the qubit excitation probability remains relatively low for the considered time interval, $%
P_{e}\lesssim 0.2$. As expected, the
cavity field becomes entangled with the qubit (as attested by nonzero
negativity), and its linear entropy increases with time. For example, for
times $\nu t\lesssim 3\times 10^{4}$, the purities of the field states are $%
1-S_{L}\gtrsim 0.7$.

\begin{figure}[tbh]
\begin{center}
\includegraphics[width=0.48\textwidth]{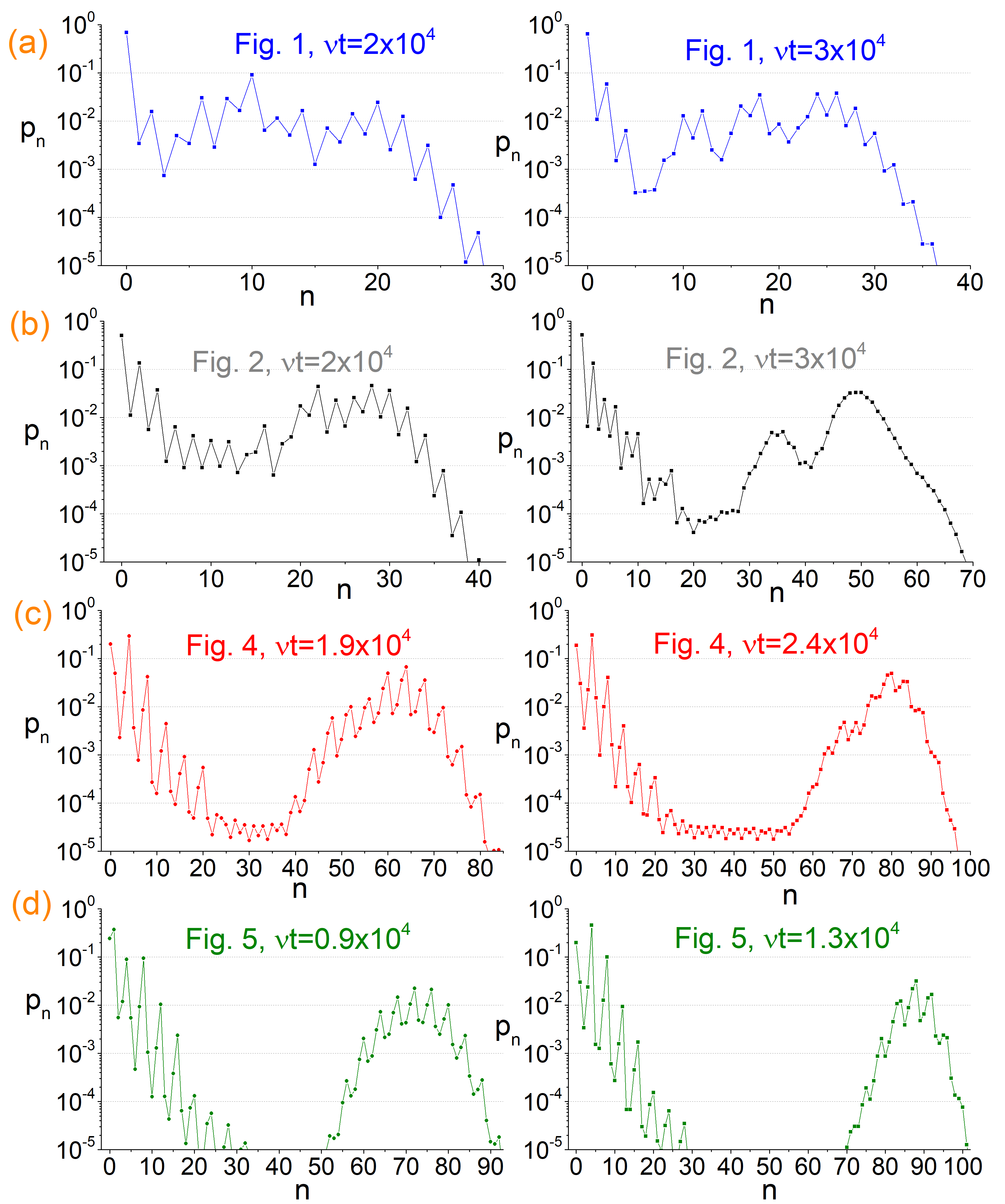} {}
\end{center}
\caption{Photon number distribution (in the log-scale) for the time instants
indicated in the plots. For the parameters of: a) Fig. \protect\ref{fig1};
b) Fig. \protect\ref{fig2}; c) Fig. \protect\ref{fig4}. d) Fig. \protect\ref%
{fig5}.}
\label{fig3}
\end{figure}

\begin{figure}[tbh]
\begin{center}
\includegraphics[width=0.48\textwidth]{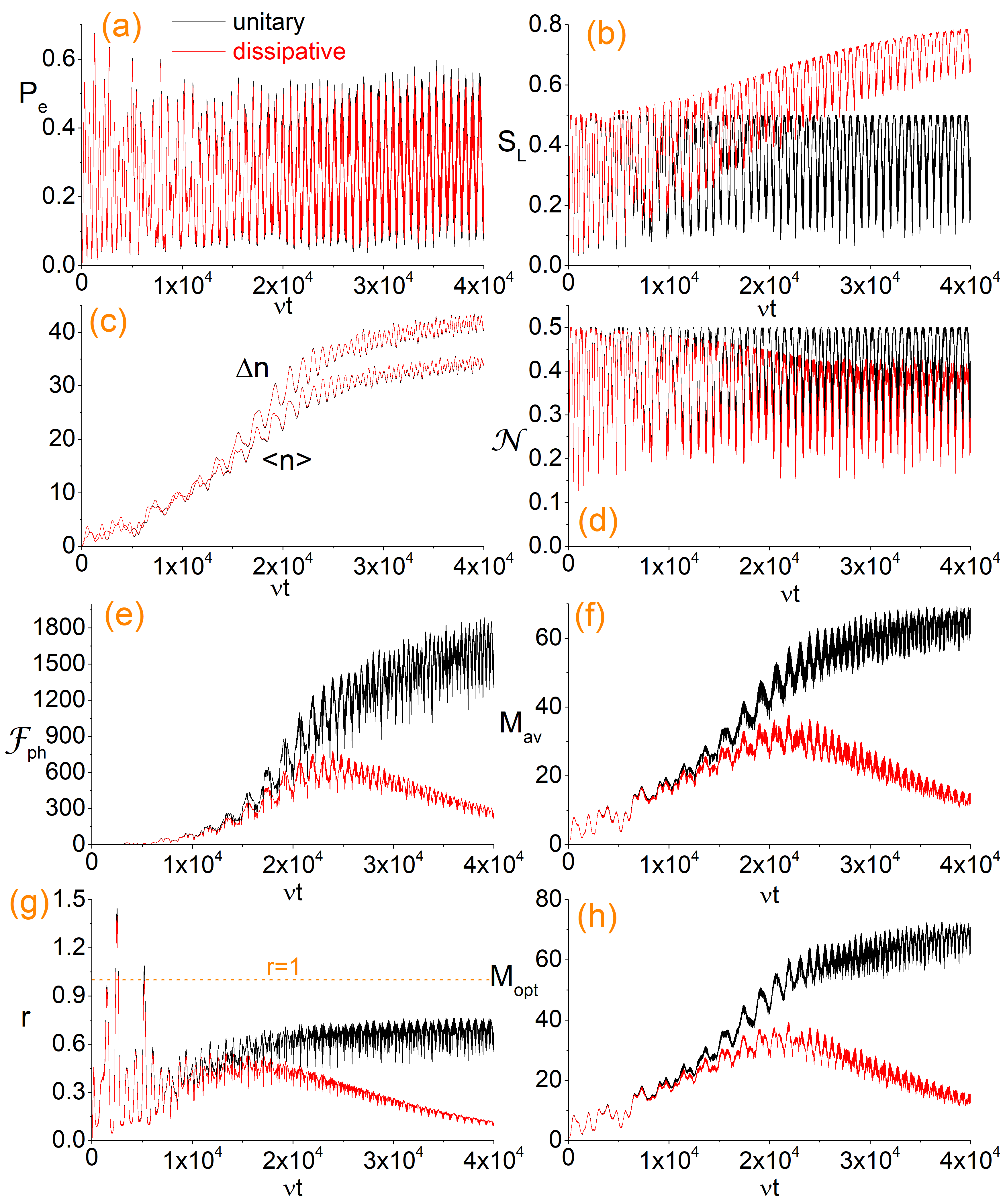} {}
\end{center}
\caption{Similar to Fig. \protect\ref{fig1}, but for the 4-photon DCE with
parameters: $g=0.15\protect\nu $, $\Omega _{0}=2.9\protect\nu $ and $\protect%
\eta _{0}=3.931\protect\nu $, $\protect\alpha =8\times 10^{-7}\protect\nu %
^{2}$.}
\label{fig4}
\end{figure}

The main results of this work, within the context of quantum metrology, are shown in panels e--h of Figs. \ref{fig1}–\ref{fig4}. In the panels %
\ref{fig1}e and \ref{fig2}e, we see that the QFI for phase estimation
exceeds significantly the average photon number, attesting that the
generated (mixed) cavity field states indeed present metrological power.
While for the considered dissipative rates the dynamics of $P_{e}$, $%
\left\langle n\right\rangle $ and $\Delta n$ are practically unaffected by
the dissipation, this is not true for the linear entropy, negativity and QFI, emphasizing the importance of taking the losses into account. Nonetheless, the metrological power
persists even in the presence of dissipation. From the panels \ref{fig1}g and \ref{fig2}g,
we see that the ratio $r$ can reach values greater than 1, without
dissipation, or of the order of $0.6$ with dissipation, meaning that the produced mixed cavity field states possess metrological power comparable to that of the squeezed vacuum states. In panels f--h of Figs. \ref{fig1} and \ref%
{fig2} we also see that $M_{av}$ and $M_{opt}$ are significantly larger
than 1, attaining values as large as $M_{av}\approx 40$ (without
dissipation) and $M_{av}\approx 20$ (with dissipation).

To better understand the origin of the metrological power in our case, panels \ref{fig3}a and \ref{fig3}b show the photon number distribution (in logarithmic scale) of the field states at the time instants $\nu t = 2 \times 10^{4}$ and $\nu t = 3 \times 10^{4}$, for the parameters used in Figs. \ref{fig1} and \ref{fig2}. These times were chosen arbitrarily to present relatively high values of $r$ and $M_{av}$, while keeping the discrepancies between the unitary and dissipative dynamics reasonably small. We observe that the distributions are quite broad and not monotonically decreasing, so the large variance of the photon number partially explains the high values of the QFI. However, the variance alone does not determine the QFI for the considered mixed states, since the QFI is strongly affected by dissipation, whereas the standard deviation is not.

\begin{figure}[tbh]
\begin{center}
\includegraphics[width=0.48\textwidth]{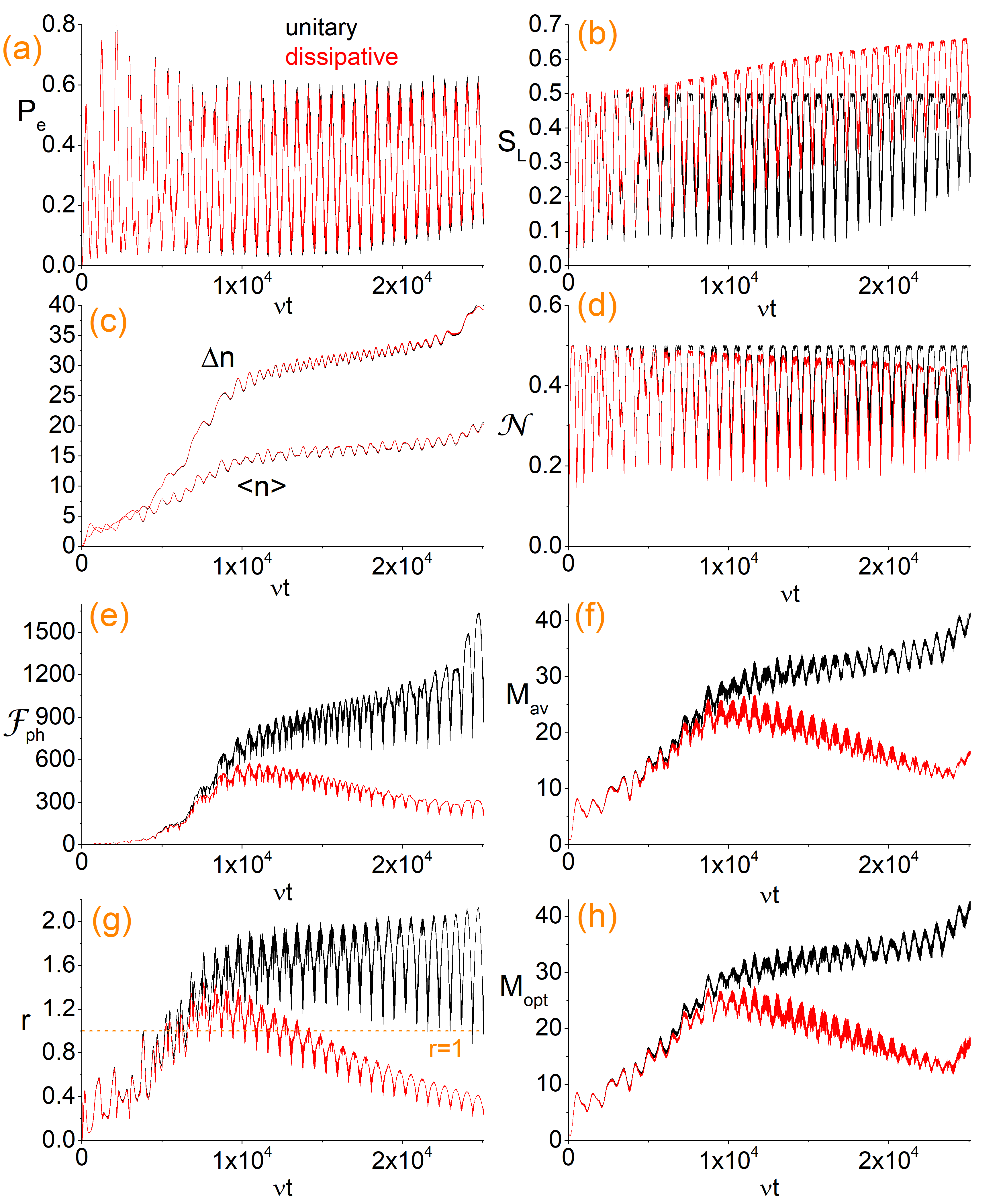} {}
\end{center}
\caption{Similar to Fig. \protect\ref{fig4}, but for the parameter $\protect%
\alpha =2\times 10^{-6}\protect\nu ^{2}$.}
\label{fig5}
\end{figure}

Finally, in Figs. \ref{fig4} and \ref{fig5}, we consider the 4-photon DCE regime \cite{AVD19}, $\eta \approx 4\nu$, near the three-photon resonance of the Rabi Hamiltonian \cite{3f1,3f2}. We set $g = 0.15\nu$ (in the ultrastrong coupling regime), $\Omega_0 = 2.9\nu$, $\eta_0 = 3.931\nu$, and consider $\alpha = 8 \times 10^{-7} \nu^2$ (Fig. \ref{fig3}) and $\alpha = 2 \times 10^{-6} \nu^2$ (Fig. \ref{fig4}). In this regime, the qubit displays a higher probability of excitation, $P_e \gtrsim 0.5$ for certain time intervals, and the purity of the cavity field state may drop below 0.5 due to significant qubit–field entanglement. The QFI is significantly larger than the average photon number, while the ratio $r$ can exceed 1 (see Fig. \ref{fig5}g), indicating that the generated states may possess more metrological power than squeezed vacuum states with the same average photon number. For the considered time interval, $M_{av}$ and $M_{opt}$ also reach high values, on the order of $M_{av} \approx 60$ (without dissipation) and $M_{av} \approx 30$ (with dissipation). Panels \ref{fig3}c and \ref{fig3}d show the photon number distribution of the cavity field states at selected time instants. We observe that the photon number distribution exhibits local peaks when $n/4$ is an integer, has a large variance, and resembles a bimodal distribution for the times considered. Note that an analytical description of our system in the considered parameter regimes is challenging, since nearly 100 Fock states actively participate in the dynamics, and the counter-rotating terms in the Rabi Hamiltonian cannot be neglected.

\section{Conclusions\label{sec3}}

In conclusion, we numerically studied a dissipative circuit QED system in which the qubit level splitting undergoes sinusoidal external modulation with a linearly varying frequency. We considered regimes where the qubit is far detuned from the resonator and the modulation frequencies are close to the resonances corresponding to the two- and four-photon dynamical Casimir effects. We showed that, for carefully chosen parameters, it is possible to generate nonclassical states of light from the vacuum with high quantum Fisher information for phase and displacement estimation, capable of outperforming squeezed vacuum states with the same average photon number. The states generated in our scheme are intrinsically mixed and exhibit photon number distributions that are markedly different from those of standard squeezed states. Therefore, single-qubit DCE, in addition to enabling the creation of excitations from the vacuum, may also serve as a resource for producing states of light with significant metrological power.

\section{Acknowledgment}

A. P. C. acknowledges the financial support by the Brazilian agency Coordena%
\c{c}\~{a}o de Aperfei\c{c}oamento de Pessoal de N\'{\i}vel Superior (CAPES,
Finance Code~001). H. R. S. acknowledges the financial support of the
Brazilian agency Funda\c{c}\~{a}o de Apoio \`{a} Pesquisa do Distrito
Federal (FAPDF, program PIBIC). A. V. D. acknowledges a partial financial
support of the Brazilian agency Funda\c{c}\~{a}o de Apoio \`{a} Pesquisa do
Distrito Federal (FAPDF, grant number 00193-00001817/2023-43).

\end{document}